\documentclass[reprint]{JASA}

\begin{document}

\title{A unifying Rayleigh-Plesset-type equation for bubbles in viscoelastic media}

\author{Alexandros T. Oratis}
\affiliation{Physics of Fluids Group, Max Planck Center for Complex Fluid Dynamics, Faculty of Science and Technology, Technical Medical (TechMed) Center, University of Twente, Enschede, The Netherlands}

\author{Kay Dijs}
\affiliation{Physics of Fluids Group, Max Planck Center for Complex Fluid Dynamics, Faculty of Science and Technology, Technical Medical (TechMed) Center, University of Twente, Enschede, The Netherlands}

\author{Guillaume Lajoinie}
\affiliation{Physics of Fluids Group, Max Planck Center for Complex Fluid Dynamics, Faculty of Science and Technology, Technical Medical (TechMed) Center, University of Twente, Enschede, The Netherlands}

\author{Michel Versluis}
\affiliation{Physics of Fluids Group, Max Planck Center for Complex Fluid Dynamics, Faculty of Science and Technology, Technical Medical (TechMed) Center, University of Twente, Enschede, The Netherlands}

\author{Jacco H. Snoeijer}
\email[]{j.h.snoeijer@utwente.nl}
\affiliation{Physics of Fluids Group, Max Planck Center for Complex Fluid Dynamics, Faculty of Science and Technology, Technical Medical (TechMed) Center, University of Twente, Enschede, The Netherlands}





\begin{abstract}
Understanding the ultrasound pressure-driven dynamics of microbubbles confined in viscoelastic materials is relevant for multiple biomedical applications, ranging from contrast-enhanced ultrasound imaging to ultrasound-assisted drug delivery. 
The volumetric oscillations of spherical bubbles is analyzed using the Rayleigh-Plesset equation, which describes the conservation of mass and momentum in the surrounding medium. 
Several studies have considered an extension of the Rayleigh-Plesset equation for bubbles embedded into viscoelastic media, but these are restricted to a particular choice of constitutive model and/or to small deformations. 
Here, we derive a unifying equation applicable to bubbles in viscoelastic media with arbitrary complex moduli and that can account for large bubble deformations. To derive this equation, we borrow concepts from finite-strain theory. 
We validate our approach by comparing the result of our model to previously published results and extend it to show how microbubbles behave in arbitrary viscoelastic materials. 
In particular, we use our viscoelastic Rayleigh-Plesset model to compute the bubble dynamics in benchmarked viscoelastic liquids and solids.  
\end{abstract}


\maketitle

\section{Introduction}
The growth and collapse of bubbles in viscoelastic materials is a physical process that features in a variety of environmental, food processing and industrial settings.
From the bubbles formed during volcanic eruptions \citep{ichihara2008dynamics} to those used in the production of polymer foams \citep{everitt2003bubble,andrieux2018liquid}, understanding the influence of viscoelasticity on the bubble dynamics is a vital step for technological innovation.
In particular, the crucial importance of viscoelasticity has been a topic of focus for biomedical applications \citep{yang2005model,gaudron2015bubble,dollet2019bubble}.
Indeed, micrometer-size bubbles are frequently used in ultrasound as contrast agents to visualize organ perfusion and blood flow down to the smallest capillary vessels \citep{frinking2020three}. 
They are also heavily investigated as therapeutic agents for cancer treatment, blood-brain-barrier opening \citep{choi2011noninvasive,sulheim2019therapeutic,alonso2015ultrasound,beccaria2013opening,sheikov2008effect}, sonoporation \citep{lentacker2014understanding,sirsi2012advances,deprez2021opening}, lithotripsy, histotripsy \citep{xu2021histotripsy}, or even sonothrombolysis \citep{miller2018ultrasonic,de2014properties,bader2016sonothrombolysis}. 
These therapeutic applications often require inducing controlled microbubble oscillations within tissue to impart treatment while preventing collateral damage to the surrounding healthy tissue. 
However, this embedding in a polymeric liquid or a soft solid has a dramatic effect on bubble dynamics \citep{dollet2019bubble}. 
More specifically these materials have viscoelastic properties and provide solid-like or liquid-like resistance to bubble motion, depending on the deformation time scale. 
Therefore, the ability to control and utilize bubbles in biomedical settings crucially relies on our understanding of the micro and macro-rheology of the surrounding medium and its impact on bubble oscillations.

Various biological materials exhibit viscoelastic properties that come with intricate behavior of stress and strain.
A simple method to understand such responses consists of modelling the viscoelastic material as a combination of Hookean springs and Newtonian dashpots \citep{bland2016theory,bird1987dynamics,kelly2013solid}.
The spring introduces elastic characteristics to the material's response, while the dashpot adds a viscous contribution. Models built from a combination of springs and dashpots typically lead to exponential relaxation of stress or strain. Even though the spring-dashpot approach accurately describes various viscoelastic liquids and solids, it fails to capture the rheology of multi-scale materials such as gels or elastomeric networks \citep{WinterChambon1986,LongAjdari1996,KarpitschkaNatureComm2015}, which often come along with power-law stress relaxation rather than exponentials. 
An alternative and more general approach consists of characterizing the material via its complex modulus $\mu(\omega) = G'(\omega) + iG''(\omega)$, which combines an elastic storage modulus $G'(\omega)$ and viscous loss modulus $G''(\omega)$. 
These moduli can be experimentally obtained by measuring the deformation of the material in response to a periodic excitation, and are functions of the imposed angular frequency $\omega$.
As an example, the complex modulus of gels typically exhibits a non-integer scaling behavior with respect to frequency, which cannot be represented by (a finite number of) springs and dashpots. 
An equivalent formulation of the rheology in the temporal domain can be achieved via the stress relaxation function $\psi(t)$. This function
describes how the stress relaxes after a material is subjected to a step-strain, and can be related to the complex modulus as $\mu(\omega) = i\omega \int_0^\infty \mathrm{d}t~ \psi(t) e^{-i\omega t}$~\citep{carcione2019kramers}; hence, $\mu(\omega)$ and $\psi(t)$ contain the exact same rheological information. 
Within this framework, $\psi \to 0$ at large times for viscoelastic liquids, reflecting how the stress fully relaxes after stopping the flow. 
Viscoelastic solids, by contrast, exhibit a finite $\psi$ at large times, and thus a residual elastic stress, expressing a long-term memory of the reference configuration at rest. 


The dynamical behavior of spherical bubbles is typically described using the Rayleigh-Plesset equation, initially developed for a spherical bubbles oscillating in free-field.
Possible ways of adding the effects of the viscoelastic surrounding on the bubble dynamics in the context of the Rayleigh-Plesset equation have been the focus of many studies, starting with the pioneering work of Fogler and Goddard \citep{fogler1970collapse,fogler1971oscillations}. 
Assuming a linear Maxwell model, the authors investigated the effects of elasticity in the cavitation of a spherical void.
Subsequent studies extended this approach by considering different constitutive models of the surrounding material~\citep{tanasawa1970dynamic,shima1986nonlinear,allen2000dynamicsA,yang2005model,cunha2013oscillatory,hamaguchi2015linear}, including those applicable to large-amplitude bubble deformations~\citep{ting1975viscoelastic,papanastasiou1984bubble,kim1994collapse,allen2000dynamicsB,jimenez2005bubble,naude2008periodic,gaudron2015bubble,warnez2015numerical}. Yet, a unifying equation valid for large deformations and applicable to materials with arbitrary complex rheology (equivalently, arbitrary $\psi(t)$), is still lacking.

In this paper we derive a Rayleigh-Plesset-type equation for arbitrary viscoelastic materials, i.e. materials described by an arbitrary stress relaxation function $\psi(t)$, that is also valid for large bubble deformations. 
For this, we resort to a modelling framework that combines linear relaxation and finite strain. 
In the context of viscoelastic solids this class of models is referred to as finite linear viscoelasticity \citep{wineman2009nonlinear}, while for viscoelastic liquids this corresponds to the K-BKZ model (Kaye-Bernstein, Kearsley, Zapas) \citep{tanner1988bk,mitsoulis201350}.
Special cases of this modelling framework include the neo-Hookean solid and the Oldroyd-B fluid.
An overview of commonly used constitutive models is provided in Tab.\,\ref{tab:constitutive}. 
As is known in viscoelasticity, there is an ambiguity in the choice of upper-convected or lower-convected tensor quantities, representing how tensors are transported by the flow. We primarily focus on the more common upper-convected models, for which, as we will demonstrate, the viscoelastic Rayleigh-Plesset equation for a bubble with radius $R(t)$ takes the form:

\begin{equation}
\begin{split}
    & \rho \left(R \Ddot{R} + \frac{3}{2}\dot{R}^2\right)  = \Delta p - \frac{2\gamma}{R}\\
     &- 2\int_{-\infty}^{t}\mathrm{d}t'\,\psi(t-t')\,\frac{\dot{R}(t')}{R(t)}\left\{  
     \left[\frac{R(t')}{R(t)} \right]^3 +   1 \right\}.
\end{split}
         \label{eq:RP_VE}
\end{equation}
Here, the dot denotes a differentiation with time, $\rho$ is the density of the medium, $\gamma$ the interfacial tension, and $\Delta p = p_g - p_\infty$ the difference between the gas pressure $p_g$ and the far field pressure $p_\infty$. 
Below, we will provide the detailed derivation of Eq.~\eqref{eq:RP_VE}, as well as some benchmarks with existing literature.
Finally, as an example of specific relevance, we will explore the resonance behavior of microbubbles (with and without coating of phospholipid molecules) oscillating in viscoelastic media.

\begin{table*}[t!]
  \centering
  \begin{tabular}{l  c c  c c}
    \hline
    \hline
    \textbf{Model} & \textbf{Schematic} & \textbf{Relaxation Function}&  \textbf{Complex Modulus} & \textbf{Constitutive} \\
        & & $\psi(t)$ &  $\mu(\omega) = G'(\omega) +i G''(\omega)$ & \textbf{Differential Equation}\\ \hline
    Viscous Fluid
    &
    \begin{minipage}{0.135\textwidth}
      \includegraphics[width=\linewidth]{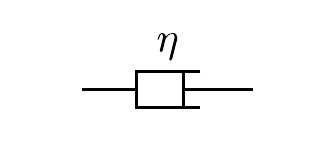}
    \end{minipage}
    & 
    $\eta \delta (t)$
    &
    $i \eta \omega$
    &$\boldsymbol{\tau} = \eta \dot{\boldsymbol{\epsilon}}$
    \\
    Maxwell Fluid
    &
    \begin{minipage}{0.135\textwidth}
      \includegraphics[width=\linewidth]{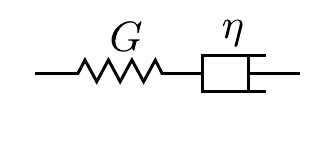}
    \end{minipage}
    & 
    $G e^{-t/\lambda}$
    &
    $\frac{i \eta \omega}{1+i \lambda\omega}$
    &$\boldsymbol\tau +\lambda\overset{\triangledown}{\boldsymbol{\tau}} = \eta \dot{\boldsymbol\epsilon}$
    \\     
    Oldroyd-B Fluid
    &
    \begin{minipage}{0.135\textwidth}
      \includegraphics[width=\linewidth]{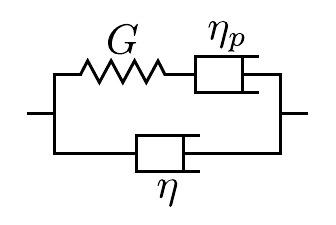}
    \end{minipage}
     & 
 $\eta \delta(t) +  Ge^{-t/\lambda}$
    &
        $\frac{-\eta \lambda \omega^2 + i \omega (\eta+ \eta_p)}{1+i\lambda\omega}$
 &$\boldsymbol\tau +\lambda\overset{\triangledown}{\boldsymbol{\tau}} = (\eta+\eta_p)\dot{\boldsymbol{\epsilon}}+\eta\lambda\overset{\triangledown}{\dot{\boldsymbol{\epsilon}}}$ 
    \\    
    Critical Gel & -
        & $G(\lambda/t)^{1/2}$ 
    & 
	$G\left(i \pi \lambda\omega \right)^{1/2}$    
    & Not available
    \\
    \hline
    Neo-Hookean Solid
    &
    \begin{minipage}{0.135\textwidth}
      \includegraphics[width=\linewidth]{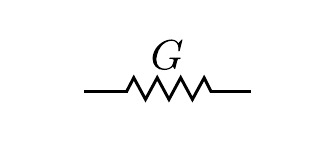}
    \end{minipage}
        & 
    $G$
    &
    $G$
    & $\boldsymbol \tau = G  \left(\mathrm{\textbf{B}} -\mathbf I \right)$
    \\
    Kelvin-Voigt Solid
    &
    \begin{minipage}{0.135\textwidth}
      \includegraphics[width=\linewidth]{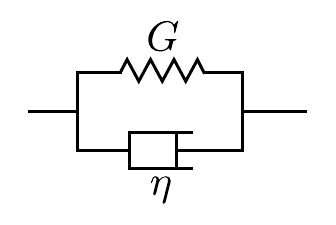}
    \end{minipage}
        & 
    $G + \eta \delta(t)$
    &
    $G + i\eta\omega$
    & $\boldsymbol \tau= G  \left(\mathrm{\textbf{B}} -\mathbf I\right) +\eta \dot{\boldsymbol \epsilon}$
    \\    
    Chasset-Thirion & - & $G[1+\Gamma(1-n)^{-1}\left(\frac{\lambda}{t}\right)^{n}]$ & $G\left[1+(i\lambda\omega)^n \right]$ & Not available \\
    \hline
    \hline
  \end{tabular}
  \caption{Examples of viscoelastic fluid and solid models, defined by the stress relaxation function $\psi(t)$ and complex modulus $\mu(\omega)$, that are captured by Eq.~(\ref{eq:RP_VE}). In some (but not all) cases the integral form of the constitutive law (see Eq.~\ref{eq:KBKZ} below) can be written as a differential equation (constitutive DE), involving upper convected derivatives, the rate-of-strain tensor $\dot{\mathbf \epsilon}$, and for solids involves the Finger tensor $\mathbf B$. 
  Note that the relaxation function of the Chasset-Thirion model contains the gamma function $
  \Gamma(t) = \int_0^\infty x^{t-1} e^{-x} \mathrm{d}x$.
}\label{tab:constitutive}
\end{table*}

\section{Derivation of the viscoelastic Rayleigh-Plesset equation}
\subsection{Constitutive equation}
We start by considering a spherical bubble, whose  radius $R(t)$ varies only in time.
We employ a spherical coordinate system $(\hat{\mathbf{e}}_r,\hat{\mathbf{e}}_\theta,\hat{\mathbf{e}}_\phi)$ at the center of the bubble, with $r$, $\theta$, and $\phi$ denoting the radial, azimuthal, and polar directions, respectively.
Assuming a purely radial incompressible flow, the flow velocity writes $\mathbf{v} = (\dot{R}R^2/r^2)\,\mathbf{\hat{e}}_r$.
The momentum equation combined with the appropriate boundary conditions can be used to obtain the Rayleigh-Plesset equation \citep{prosperetti1982generalization}
\begin{equation}
\begin{split}
    &\rho \left(R\ddot{R} + \frac{3}{2}\dot{R}^2\right) = \Delta p - \frac{2\gamma}{R}+\\
    &+ \int_{R}^{\infty}\mathrm{d}r\,\frac{1}{r} (2\tau_{rr} - \tau_{\theta\theta}- \tau_{\phi\phi}),
\end{split}
    \label{eq:RP}
\end{equation}
where $\tau_{rr}$, $\tau_{\theta\theta}$, and $\tau_{\phi\phi}$ represent the  radial, azimuthal and polar components of the deviatoric stress tensor $\boldsymbol{\tau}$.
In the context of small bubble deformations, the stress tensor is usually assumed to be traceless, such that  $2\tau_{rr} - \tau_{\theta\theta}- \tau_{\phi\phi} = 3\tau_{rr}$.
This assumption, often used to simplify the analysis, allows for an exact integration for specific constitutive laws (e.g. Kelvin-Voigt model \citep{yang2005model,hamaguchi2015linear}).
This strategy, however, is not suited to establish a generalized Rayleigh-Plesset equation that encompasses arbitrary viscoelastic models for large deformations, which is the present goal of this paper. 
Instead, we must maintain the integral as in Eq.~\eqref{eq:RP} and proceed with the so-called finite linear viscoelastic formulation \citep{wineman2009nonlinear}.
The essence of finite linear viscoelasticity is to combine a linear relaxation in time, while admitting geometric nonlinearities associated to large deformations. 
The corresponding constitutive relation involves an integral over the history of deformation, and is of the form:
\begin{equation}
\begin{split}
    \boldsymbol\tau = - \int_{-\infty}^t \mathrm{d}t'&\, [\psi_1(t-t') \frac{\partial \mathbf{B}(t,t')}{\partial t'}\\ 
    &- \psi_2(t-t') \frac{\partial \mathbf{B}^{-1}(t,t')}{\partial t'}].
\end{split}
    \label{eq:KBKZfull}
\end{equation}
Here, we have introduced the Finger tensor $\mathbf B(t,t')$, which expresses the deformation between the states at time $t'$ in the past and the current time $t$. 
A precise definition of $\mathbf B(t,t')$ is given below. Equation~(\ref{eq:KBKZfull}) resembles the K-BKZ model for viscoelastic liquids \citep{tanner1988bk,mitsoulis201350}, which can be recovered via an integration by parts. 
The appearance of two relaxations functions, respectively associated to $\mathbf B$ and to its inverse $\mathbf B^{-1}$, reflects the freedom of choosing upper-convected or lower-convected derivatives of tensors in viscoelastic  models (see Appendix~\ref{app:UCM}). 
We will focus on upper-convected materials, which are based on  $\mathbf B$ rather than its inverse \citep{snoeijer2020relationship}. Therefore, setting $\psi_1=\psi$ and $\psi_2=0$, we obtain: 
\begin{equation}
    \boldsymbol\tau = - \int_{-\infty}^t \mathrm{d}t'\,\psi(t-t') \frac{\partial \mathbf{B}(t,t')}{\partial t'} ,
    \label{eq:KBKZ}
\end{equation}
which is the constitutive relation used in the remainder of this paper. 
For completeness, the result obtained from using the lower convective derivative is worked out in Appendix~\ref{app:Lower}.

Two remarks are in order here. First, in the limit of small deformations where $t'\to t$, the time derivative of the Finger tensor reduces to $\frac{\partial \mathbf{B}(t,t')}{\partial t'} = - \dot{\boldsymbol{\epsilon}}(t)$, where $\dot{\boldsymbol{\epsilon}}(t)= \nabla \textbf{v} + (\nabla \textbf{v})^{\mathrm{T}}$ is the rate of strain tensor (Appendix~\ref{app:Kinematics}, see also \citep{essink2022soft}). 
Inserting this expression into Eq.~\eqref{eq:KBKZ}, we obtain the conventional memory integral for the stress at small deformations~\citep{bird1987dynamics}, as is frequently used in the context of bubbles \citep{fogler1970collapse,dollet2019bubble}. 
In general, however, $\frac{\partial \mathbf{B}(t,t')}{\partial t'}$ is not equal to $\dot{\boldsymbol \epsilon}$, and this distinction is essential at large deformations. 
Second, (\ref{eq:KBKZ}) reflects the deformation history of a certain material point. By consequence, the integral must be carried out at constant material point and calls for a Lagrangian description of the problem.

\subsection{Lagrangian formulation}
To define the Finger tensor $\textbf{B}(t,t')$ we introduce a Lagrangian description of the deformation, which features prominently in finite-strain theory.
Specifically, this description involves relating the Eulerian position $\mathbf{x} = \xi(\mathbf{X},t)$ in the current configuration at time $t$ to a Lagrangian material point $\mathbf{X}$. 
In solids, one naturally defines $\mathbf{X}$ as the coordinates in the reference configuration, but, in general, one can define $\mathbf X$ from the configuration  at some arbitrary time $t_0$. 
The mapping between the two states $\xi$ can be used to evaluate the deformation gradient tensor $\mathbf{F}(t,t_0) = \frac{\partial \mathbf{x}}{\partial \mathbf{X}}$. 
This tensor describes the material stretching through the transformation of a line element $d\mathbf X$ (i.e. the distance between two material particles) in the reference configuration at time $t_0$ to the same material line element $d\mathbf x$ at time $t$. 
Using the deformation gradient, the Finger tensor is defined as $\mathbf{B}(t,t_0) = \mathbf{F}(t,t_0) \cdot \mathbf{F}^{\mathrm{T}}(t,t_0)$.

\begin{figure}[t]
    \centering
    \includegraphics[width=0.999\linewidth]{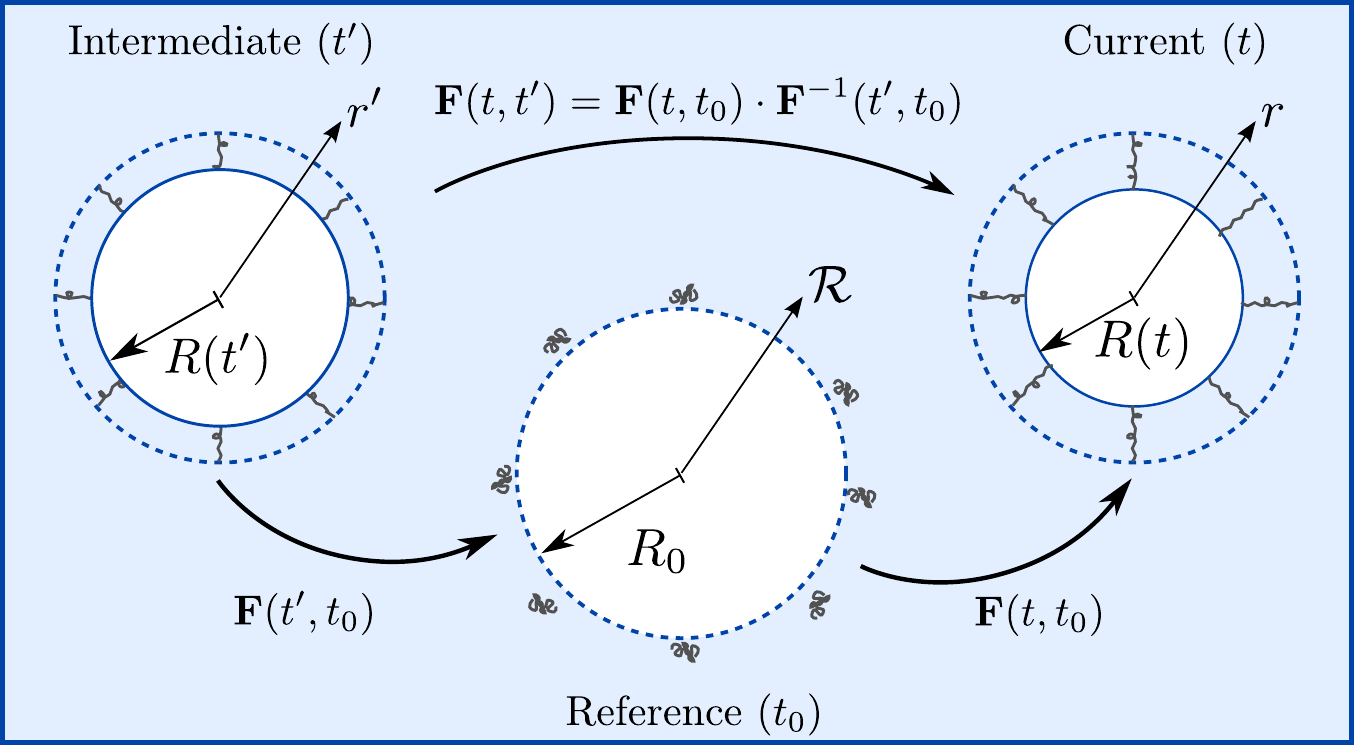}
    \caption{Schematic illustrating the mapping between the reference and current state for a bubble deforming with radius $R(t)$.
    The material coordinate $r$ in the current state can be related to the undeformed material coordinate $\mathcal{R}$.
    To evaluate the stress in terms of the history of deformation, it is instructive to introduce an intermediate past state at time $t'$. 
    A new mapping can then be derived in terms of the material coordinate in the intermediate state $r'$.
    The mapping between each state can be used to determine the deformation gradient tensor $\mathbf{F}$.}
    \label{fig:mapping}
\end{figure}

During the purely spherical motion of the bubble, we only need to keep track of the radial position of the materials points.
In Fig.\,\ref{fig:mapping} we therefore denote $\mathcal R$ as the reference radial position of any point in the medium.
On the interface of the bubble at rest, $\mathcal R = R_0 = R(t_0)$, which is thus the bubble radius in the reference configuration. 
Since the deformation is unidirectional, the tensor $\mathbf F$ is diagonal: the radial component of the deformation gradient tensor reads $\frac{\partial r}{\partial \mathcal R}$, while the azimuthal components are given by the ratio $\frac{r}{\mathcal R}$ \cite{holzapfel2002nonlinear}. 
The latter represents the stretching of a shell of constant material point. 
We thus find

\begin{equation}
\mathbf{F}(t,t_0) = 
\begin{pmatrix}
	\dfrac{\partial r}{\partial \mathcal R} & 0 & 0 \\
	0 & \dfrac{ r}{ \mathcal R} & 0 \\
	0 & 0 & \dfrac{ r}{ \mathcal R}.
\end{pmatrix}
\end{equation}
Incompressibility of the medium requires that $\mathrm{det} (\mathbf{F}) = 1$ \cite{holzapfel2002nonlinear}, which implies $\frac{\partial r}{\partial \mathcal R} = (\frac{ \mathcal{R}}{r})^2$. Integrating this relation, we identify the mapping between the two states: 

\begin{equation}\label{eq:radialmapping}
r^3(t,\mathcal R) = \mathcal{R}^3 + R(t)^3 - R_0^3,
\end{equation}
where the time-dependence is entirely encoded in the difference between the bubble radius $R(t)$ compared to the reference radius $R_0$. This mapping has indeed been used in many previous studies \citep{fogler1970collapse,fogler1971oscillations,ting1975viscoelastic,papanastasiou1984bubble,allen2000dynamicsA,allen2000dynamicsB,gaudron2015bubble}. 
For this spherical geometry, we simply recover the conservation of volume for concentric spheres: $r^3-R^3 = \mathcal R^3 -R_0^3$. 
\cite{allen2000dynamicsB}

To further evaluate the memory integral in Eq.\,\eqref{eq:KBKZ} we need to express the deformation in terms of the entire history of the bubble motion, and not with respect to the reference state. We thus introduce the position $\mathbf x'$ as the position of a material point at some past time $t' < t$.
The deformation gradient between two arbitrary times $t$ and $t'$ then becomes $\mathbf{F}(t,t') = \frac{\partial \mathbf{x}}{\partial \mathbf{x'}} = \mathbf{F}(t,t_0) \cdot \mathbf{F}^{-1}(t',t_0)$. As explained schematically in Fig.\,\ref{fig:mapping}, the mapping between $t'$ and $t$ can thus be obtained in two steps: first moving from the configuration at $t'$ to the reference configuration at $t_0$, and then going from $t_0$ to $t$. Bearing in mind that $\mathbf B(t,t') = \mathbf F(t,t')\cdot \mathbf F(t,t')^\mathrm{T}$, we thus obtain the components of the Finger tensor as 
\begin{equation}
\begin{split}
B_{\phi\phi}(t,t') = B_{\theta \theta}(t,t') &= \left(\frac{r}{r'}\right)^2\\
&=
\left(\frac{\mathcal R^3 + R(t)^3 - R_0^3 }{\mathcal R^3 + R(t')^3 - R_0^3}\right)^{2/3},
\end{split}
\end{equation}
\begin{equation}
\begin{split}
B_{rr}(t,t') = \frac{1}{B_{\theta \theta}(t,t')^2} &= \left(\frac{r'}{r}\right)^4 \\
&=
\left(\frac{\mathcal R^3 + R(t')^3 - R_0^3 }{\mathcal R^3 + R(t)^3 - R_0^3}\right)^{4/3}. 
\end{split}
\end{equation}
In the final step, we have made use of the explicit radial mapping of Eq.~\eqref{eq:radialmapping}. This step is crucial, as it allows expressing $\mathbf B(t,t')$ at a constant material point $\mathcal R$, as is required for the evaluation of the integral in Eq. \eqref{eq:KBKZ}. 
The components of the deviatoric stress tensor then follow as
\begin{equation}
\begin{split}
    \tau_{rr} = -4\int_{-\infty}^{t}\mathrm{d}t'\,&\{\psi(t-t')R^2(t')\,\dot{R}(t')\\
     & \frac{\left[\mathcal{R}^3+R^3(t')-R_0^3\right]^{1/3}}{\left[\mathcal{R}^3+R^3(t)-R_0^3\right]^{4/3}}\},
\end{split}
    \label{eq:tau_rr}
\end{equation}
\begin{equation}
\begin{split}
    \tau_{\theta\theta} = \tau_{\phi\phi} = 2\int_{-\infty}^{t}\mathrm{d}t'\,&\{\psi(t-t')R^2(t')\,\dot{R}(t')\\
    & \frac{\left[\mathcal{R}^3+R^3(t)-R_0^3\right]^{2/3} }{\left[\mathcal{R}^3+R^3(t')-R_0^3\right]^{5/3}} \}.
\end{split}
    \label{eq:tau_phi}
\end{equation}
The resulting components of $\boldsymbol \tau$ are now explicit functions of time, encoded in $R(t)$, and the Lagrangian position $\mathcal R$. 

The remaining step is to spatially integrate the stresses in the Rayleigh-Plesset equation, as required in Eq.~\eqref{eq:RP}. These spatial integrals can be carried out explicitly, so that we are only left with the temporal memory integrals
\begin{equation}
\begin{split}
    &\int_{R}^{\infty}\mathrm{d}r\,\frac{1}{r} (2\tau_{rr} - \tau_{\theta\theta}- \tau_{\phi\phi}) = \\
    &- 2\int_{-\infty}^{t}\mathrm{d}t'\,\psi(t-t')\,\frac{\dot{R}(t')}{R(t)}\left\{  
     \left[\frac{R(t')}{R(t)} \right]^3 +   1\right\}.
\end{split}
     \label{eq:tau_integration}
\end{equation}
This concludes the derivation of the viscoelastic Rayleigh-Plesset equation for arbitrary complex modulus as presented in Eq.~\eqref{eq:RP_VE}.

\section{Applications}
\subsection{Special cases \& benchmarking}

We contextualize the derived Rayleigh-Plesset equation by considering a set of special cases for $\psi(t)$, enabling a benchmark with existing literature. The relaxation functions of various models encountered in the literature are summarized in Table~\ref{tab:constitutive}. 
Some of these can be represented schematically by a spring-dashpot, in which case it is possible to articulate the constitutive relation as a differential equation (see schematics in Table ~\ref{tab:constitutive}). 
Note that, gel-like materials that exhibit a power-law relaxation function cannot be represented by a differential constitutive law, and thus must be treated with an integral formulation. 

We first consider a Newtonian fluid of viscosity $\eta$, for which the relaxation function takes the form $\psi(t) = \eta\,\delta(t)$, where $\delta(t)$ is the Dirac delta function. 
Using the convolution of the delta function, the integral in the Rayleigh-Plesset equation reduces to $-4\eta \dot{R}/R$, which is the standard viscous contribution. 
Second, we turn to the Neo-Hookean solid, which is a purely elastic medium, obtained when the relaxation function is constant $\psi(t) = G$. 
In this case, we can carry out explicitly the memory integral in Eq.~\eqref{eq:tau_integration}. 
Applying the initial condition $R(t'\to-\infty) = R_0$, with $R_0$ the radius in the reference configuration, yields an elastic contribution to the Rayleigh-Plesset equation of the form $(G/2)[5 - 4(R_0/R) - (R_0/R)^4]$. 
This term exactly corresponds to the one previously obtained for a bubble inside a Neo-Hookean solid \citep{gaudron2015bubble,warnez2015numerical}.
Third, we consider a bubble inside an Oldroyd-B fluid, for which the relaxation function takes the form $\psi(t) = \eta\,\delta(t) + \,G\mathrm{exp}(-t/\lambda)$, where $\eta$ is the solvent viscosity, $\lambda$ the relaxation time, and $G=\eta_p/\lambda$ the polymer's elastic modulus (see also Appendix~\ref{app:UCM}). 
Here as well, the resulting Rayleigh-Plesset equation is identical to that reported by Ting, who considered the Oldroyd-B model in the context of bubble cavitation \citep{ting1975viscoelastic}.

To further validate our model and its numerical implementation, we compare it to the results obtained by Allen and Roy \citep{allen2000dynamicsB}.
Specifically, Allen and Roy considered large deformations of an acoustically driven micron-sized bubble.
The oscillatory pressure field is $p_\infty = p_0 + p_a \sin(2\pi f t)$, where $p_0$ is the ambient pressure, $p_a$ is the pressure amplitude, and $f$ the driving frequency.
The gas pressure takes the form $p_g = (p_0 + 2\gamma/R_0)(R_0/R)^{3k}$, to account for thermal dissipation through the polytropic constant $k$.
The bubble is surrounded by virtual Upper Convected Maxwell fluids (UCM) with relaxation times $\lambda =$ 0, 0.5, and 1 $\mu$s. 
The ratio between the relaxation time and the characteristic time scale of the flow is expressed by the Deborah number, which takes the values $\mathrm{De} = 2\pi f \lambda = $ 0, 1, and 2 for the three relaxation times considered.

\begin{figure}[t!]
    \centering
    \includegraphics[width=0.95\linewidth]{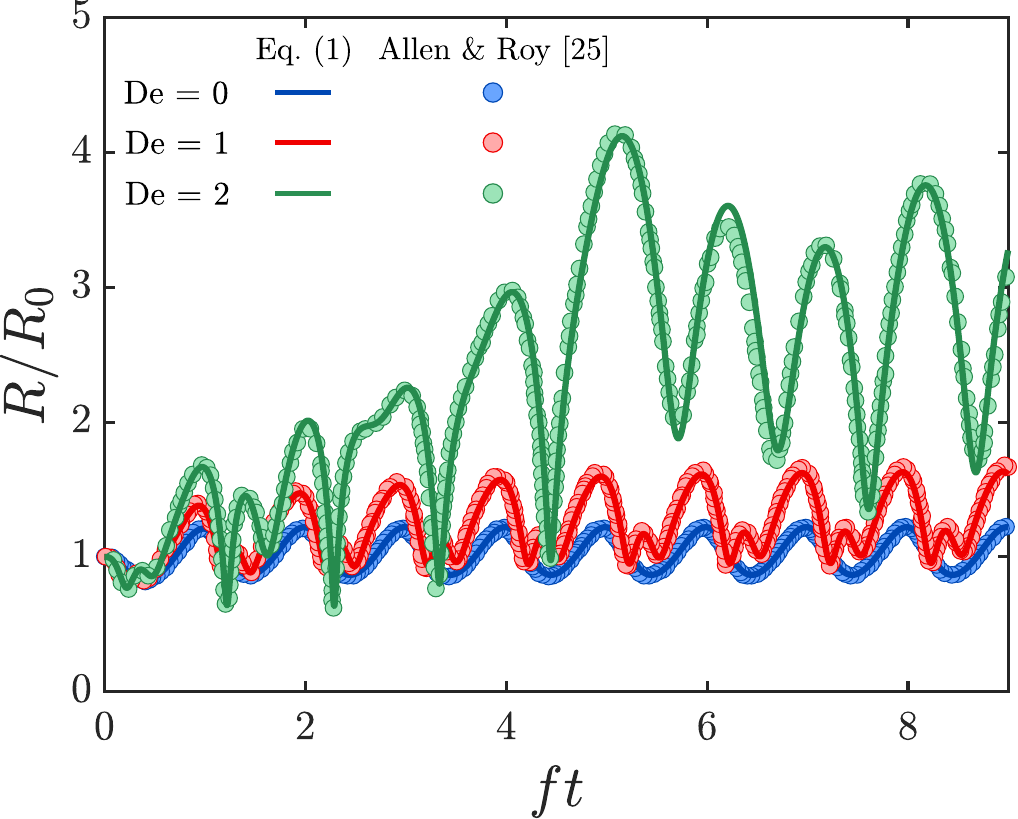}
    \caption{Plot of the normalized radius $R/R_0$ against the acoustic cycles $ft$ for a micron sized bubble in an Upper Convected Maxwell model.
    Our results using the memory integral (solid lines) are overlayed on top of those by Allen and Roy (circles) (see Fig. 12 in \citep{allen2000dynamicsB}). 
    Here we take the pressure amplitude as $p_A = 0.4$ MPa and a driving frequency $f = 3$ MHz. 
    The properties of the surrounding material are $\rho = 1000$ $\mathrm{kg/m^3}$, $\gamma = 0.072$ N/m, and $\eta = 0.03$ MPa$\cdot$s.
    The relaxation times are varied such that the Deborah number $\mathrm{De} = 2\pi f\lambda $ = 0, 1, 2 and the corresponding shear modulus is computed as $G = \eta/\lambda$. }
    \label{fig:benchmark}
\end{figure}

Our model is in excellent agreement with the results reported by Allen and Roy, as shown in Fig.~\ref{fig:benchmark}. 
Importantly, the numerical solution in \citep{allen2000dynamicsB} is not based on the integral form of the constitutive equation, but on the differential form.
Consequently, the stress field outside the bubble must be solved numerically, and the spatial integral of the stress needed in the Rayleigh-Plesset equation must also be evaluated numerically. 
Hence, the agreement in Fig.~\ref{fig:benchmark} offers a nontrivial validation of the proposed modelling framework. 
We also wish to highlight that Eq.~\eqref{eq:RP_VE} has the advantage of providing an autonomous equation for the bubble radius $R(t)$, which no longer relies on a separate (numerical) evaluation of the stress outside the bubble.

\subsection{Resonance behavior of microbubbles}

\begin{figure*}[t!]
    \centering
    \includegraphics[width=0.975\linewidth]{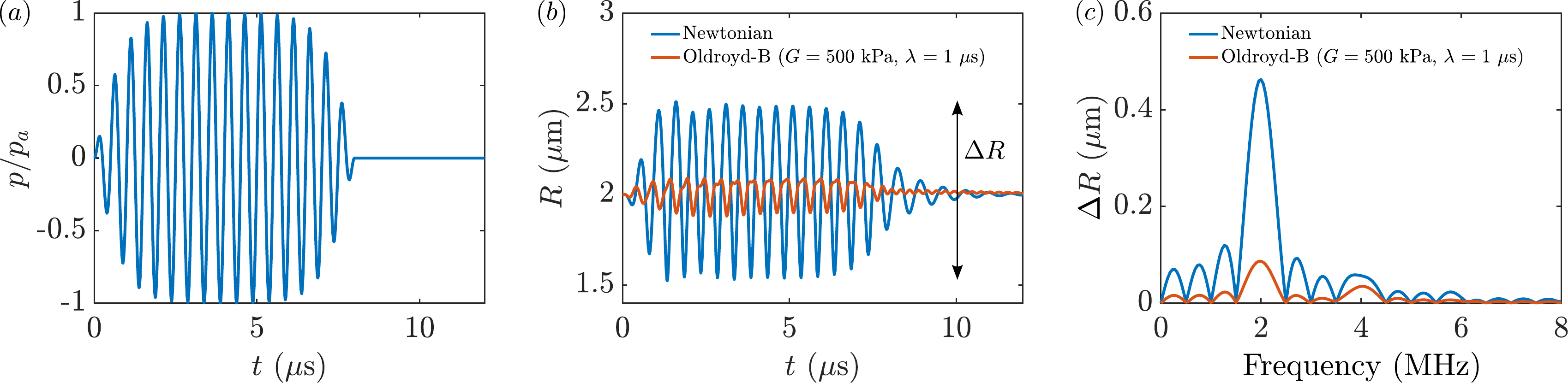}
    \caption{Response of the bubble radius $R$ when subjected to a pressure pulse $p(t)$ with frequency $f = 2$ MHz.
	(\textit{a}) The pressure pulse $p(t)$ normalized by its amplitude $p_a$ against time.    
    (\textit{b}) A bubble in a Newtonian liquid oscillates periodically following the applied pressure pulse. The bubble radius $R(t)$ achieves a maximum peak-to-peak amplitude $\Delta R \approx 0.4R_0$. 
    If the bubble is surrounded by an Oldroyd-B liquid with relaxation time $\lambda = 1$ $\mu$s and shear modulus $G = 500$ kPa, the resistance of the surrounding fluid leads to smaller oscillation amplitudes.
    (\textit{c}) Performing a fast Fourier transform of the bubble oscillation amplitude $\Delta R$, we can observe how it varies in the frequency domain, including higher and lower harmonics.}
    \label{fig:R_vs_t}
\end{figure*}

Having established the validity of our model and its numerical implementation, we now further explore the oscillatory motion of bubbles in viscoelastic materials. 
We study the motion of a 2~$\mu$m sized bubble driven by a Gaussian-tapered pressure pulse with 16 acoustic cycles (Fig.~\ref{fig:R_vs_t}a), pressure amplitude of 50 kPa and a frequency in the range $0.5\leq f\leq 8$~MHz. 
For material properties, we select a liquid density $\rho = 1,000$~$\mathrm{kg/m^3}$, and for most cases we fix the surface tension to a constant value $\gamma = 72$~mN/m.
For viscoelastic liquids we consider the bubble to be surrounded by an Oldroyd-B liquid or a Critical Gel, whose relaxation functions are $\psi(t) = G\exp(-t/\lambda)$ and $\psi(t) = G(\lambda/t)^{1/2}$ respectively (see Table~\ref{tab:constitutive}). 
For the Oldroyd-B liquid we set the solvent viscosity to $\eta = 2$~$\mathrm{mPa\cdot s}$ and we test the viscoelastic effects effects by varying the values of the shear modulus $G$ and relaxation time $\lambda$. 
Specifically, we consider three shear moduli $G = \text{10, 100, and 500}$~kPa and two relaxation times of $\lambda = \text{0.01 and 1.00}$~$\mu$s. 
The same parameters are used for the Critical Gel to enable a direct comparison. As an additional perspective, we will also consider an example where the bubble is coated by a layer of lipids, modelled via additional mechanical properties at the interface (see below). By comparing the resonance behavior of coated and uncoated bubbles, one further appreciates  relative importance of the viscoelasticity of the surrounding medium.


\subsubsection{Oldroyd-B fluid}

The results of our simulations for the evolution of the bubble radius $R(t)$ in a Newtonian and Oldroyd-B fluid are shown in Fig.~\ref{fig:R_vs_t}. 
Considering first a bubble in a Newtonian liquid, the oscillation amplitude of the bubble increases initially as it follows the applied pressure pulse. 
The bubble radius reaches stable oscillations with a peak-to-peak amplitude $\Delta R = 0.4 R_0$, before decreasing to its initial radius, as the pressure pulse decays (Fig.~\ref{fig:R_vs_t}b). 
Switching to an Oldroyd-B fluid with relaxation time $\lambda = 1$~$\mu$s and shear modulus $G = 500$~kPa, the amplitude decreases to approximately 20\% of $R_0$. 
We thus observe that the resistance of the viscoelastic stresses can significantly affect the bubble oscillation. 
To further examine the effects of the surrounding medium, it is more instructive to evaluate the resonance behavior of the bubble. 
Indeed, bubbles are known to act as damped harmonic oscillators when subjected to oscillatory pressure driving \citep{prosperetti1977thermal,minnaert1933musical}.

\begin{figure*}[t!]
    \centering
    \includegraphics[width=0.95\linewidth]{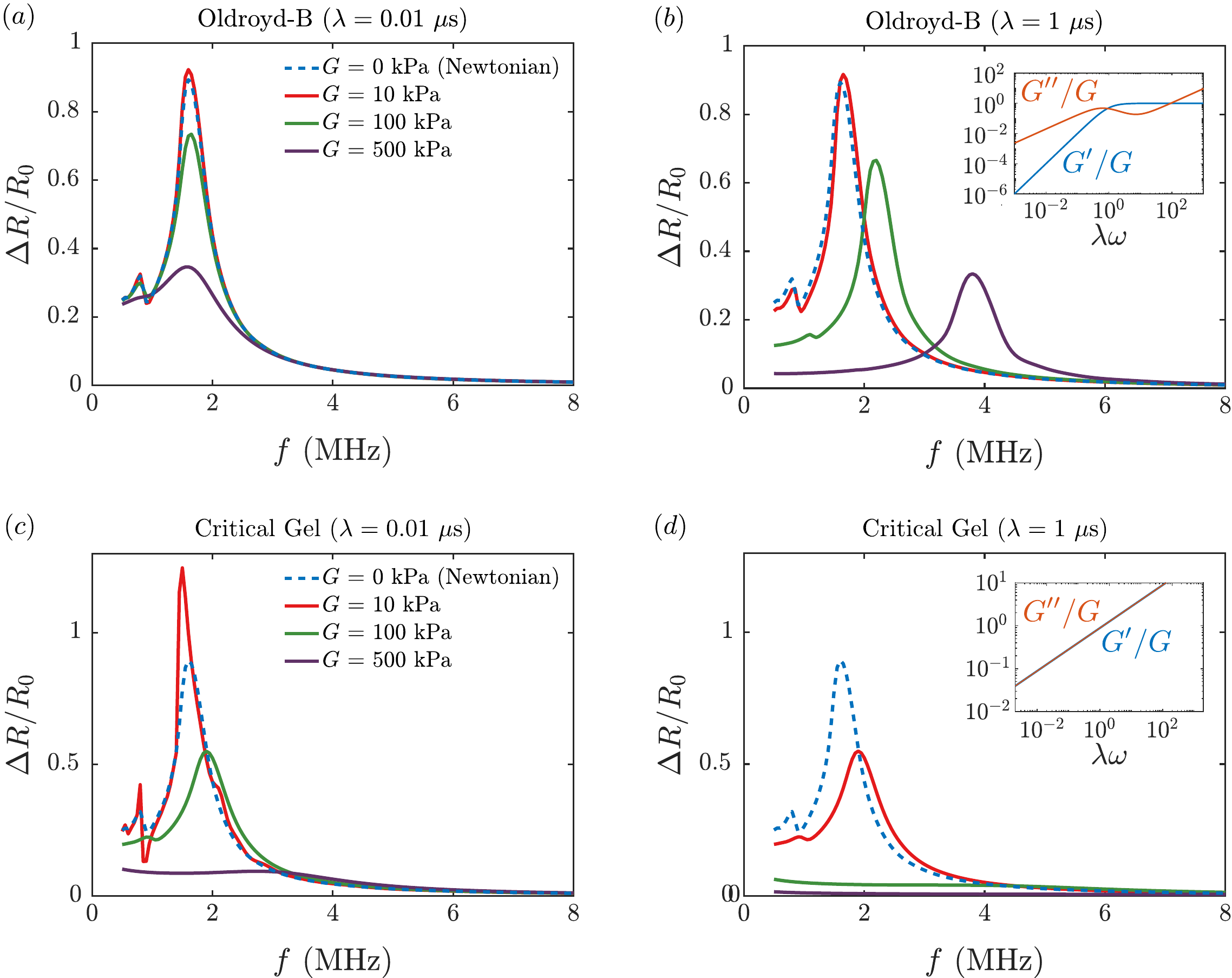}
    \caption{Effects of the shear modulus $G$ and relaxation time $\lambda$ on the bubble amplitude $\Delta R$ of a microbubble in an Oldroyd-B liquid and a Critical Gel. 
    (\textit{a}) For a relatively low relaxation time ($\lambda = 0.01$ $\mu$s) in an Oldroyd-B fluid, increasing the shear modulus leads to a decrease in the bubble amplitude. Yet, the resonance frequency remains unaffected. 
    (\textit{b}) Increasing the relaxation time to $\lambda = 1$ $\mu$s now also changes the resonance frequency of the bubble with the shear modulus. The bubble amplitude also decreases but slightly compared to the smaller relaxation time.
    Inset: The normalized storage $G'/G$ and loss $G''/G$ moduli of the Oldroyd-B fluid as a function of the normalized angular frequency $\omega \lambda$, indicating a viscous behaviour at low time scales and an elastic behavior at larger time scales. Note that the loss modulus has be computed for a viscosity ratio $\eta/\eta_p = 2\cdot 10^{-3}$.
    (\textit{c}) For a bubble in an Critical Gel with $\lambda = 0.01$ $\mu$s, both and amplitude and resonance frequency vary significantly from their Newtonian counterpart. 
    (\textit{d}) An increase in the relaxation time has much more dramatic effects in a Critical gel, where the amplitude sharply decreases for larger shear moduli.
    Inset: Because the critical gel is a special case corresponding to the gelation point, the storage and loss moduli are identical for all frequencies.}
    \label{fig:Ideal_Gel}
\end{figure*}

To this end, we obtain the response amplitude of the bubble through a fast Fourier transform and plot it as a function of the driving frequency.
The fast Fourier transform of the oscillation amplitude is also indicative of the presence of higher or lower harmonics (Fig.~\ref{fig:R_vs_t}c). 
The resulting normalized resonance curves of  $\Delta R/R_0$ are shown in Fig.~\ref{fig:Ideal_Gel}. 
For a 2~$\mu$m radius bubble in a Newtonian liquid and a polytropic constant $k=1.4$, the resonance frequency is very close to the Minnaert frequency $f_0 \approx 1/(2\pi R_0) (3kp_0/\rho)^{1/2} \approx 1.6$~MHz \cite{minnaert1933musical}, where the bubble displays a maximum relative amplitude of oscillation of 0.42 (Fig.~\ref{fig:Ideal_Gel}a). 
The viscoelastic effects become immediately apparent for the Oldroyd B liquid; with a relaxation time $\lambda = 0.01$~$\mu$s, the viscoelastic effects almost exclusively translate into a reduction of the oscillation amplitude. 
Note that $G = 10$~kPa does not give rise to significant difference as compared to the Newtonian case.
By contrast, shear moduli of 100 and 500 kPa decrease the response amplitude by 20 and 50\%, respectively. 
In all three cases the resonance frequency remains unaffected. The effects of viscoelasticity change qualitatively when increasing the relaxation time to $\lambda = 1$~$\mu$s (Fig.~\ref{fig:Ideal_Gel}b). 
Even though $G = 10$~kPa does not significantly change the bubble behavior, setting $G = 100$ and 500 kPa results in a drastic increase in resonance frequency (20 and 100 \%, respectively). 
In addition, there is a further decrease in amplitude as compared to the case of $\lambda = 0.01$~$\mu$s.

\begin{figure}[t!]
    \centering
    \includegraphics[width=0.975\linewidth]{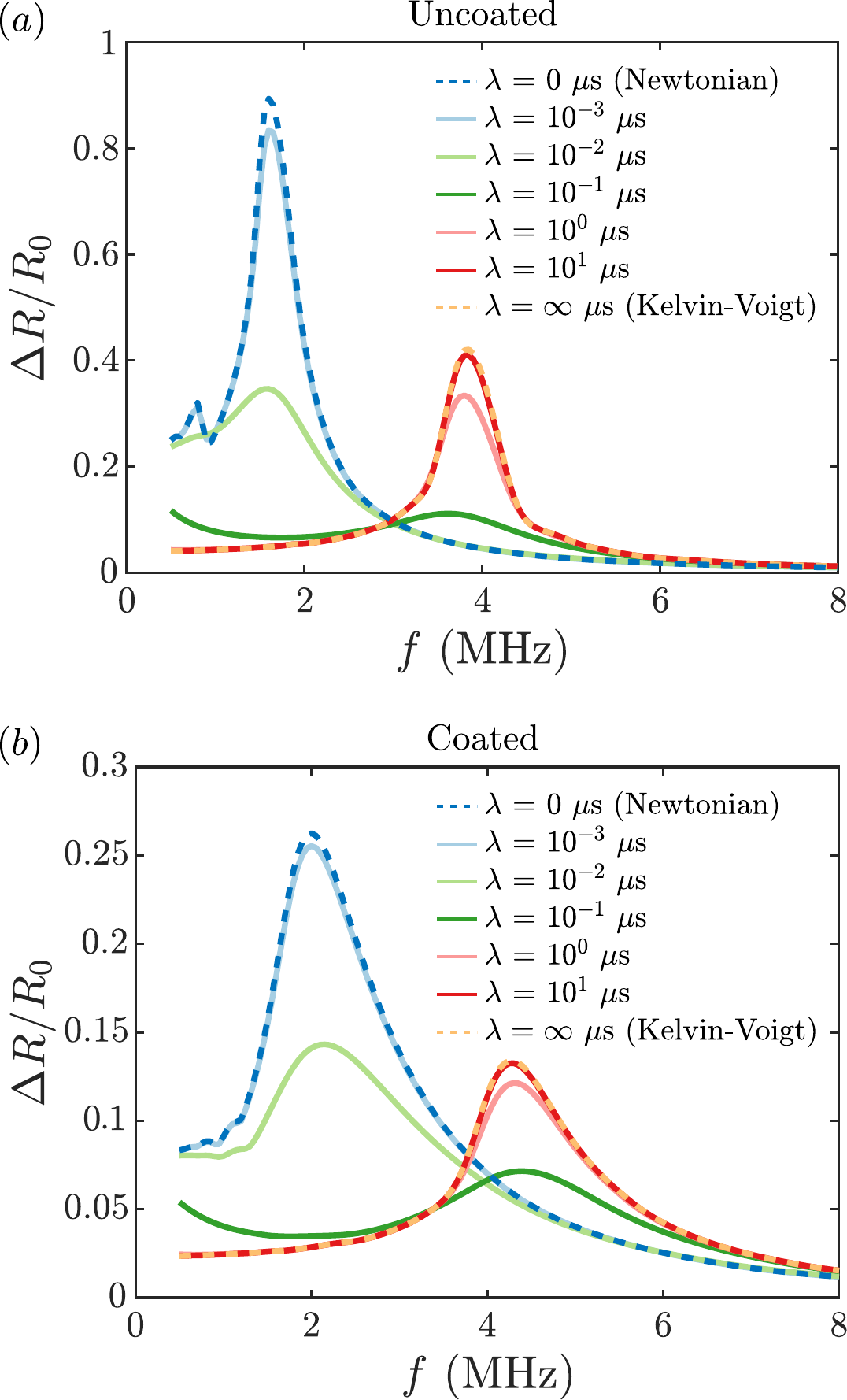}
    \caption{Convergence of the Oldroyd-B fluid to a Kelvin-Voigt solid.
    (\textit{a}) As we gradually increase the relaxation time of an Oldroyd-B fluid ($G = 500$ kPa) its behaviour starts deviating from a Newtonian liquid and converges to that of a Kelvin-Voigt solid. The amplitude at resonance follows a non-monotonic trend with the relaxation time, highlighting the different effects of viscosity and elasticity at different relaxation times.
    (\textit{b}) Considering a coated microbubble, the resonance curve qualitatively follows the same trends, as it converges to a  Kelvin-Voigt solid with an increase in the relaxation time. Quantitatively, the additional viscous and elastic resistance of the coating further decrease the oscillation amplitude and slightly increase the resonance frequency.
}
    \label{fig:Oldroyd_B}
\end{figure}

These results can be explained by calculating the Deborah number $\mathrm{De} = \lambda f$, which compares the relaxation time $\lambda$ to the characteristic time scale of the flow. 
For a relaxation time $\lambda = 0.01$~$\mu$s, the maximum Deborah number is De = 0.1, allowing the elastic stresses sufficient time to relax. 
As a result, the Oldroyd-B fluid at these low extensional rates exhibits a viscous response rather than an elastic contribution and $G''$ dominates (Fig.~\ref{fig:Ideal_Gel}b inset). 
The effective polymer viscosity $\eta_p = G\lambda$ works in conjunction with that of the solvent $\eta$ to dampen the oscillation amplitude. 
For shear moduli of 10, 100, and 500 kPa, the polymer viscosity takes a value of $\eta_p = $ 0.1, 1, and 5~mPa$\cdot$s, respectively. 
Only the latter two are high enough as compared to the solvent viscosity $\eta$ = 2 mPa·s to affect the dynamics, which is confirmed by our results. 
In contrast, for a relaxation time of $\lambda = 1$ $\mu$s the maximum Deborah number takes a value of De = 10. 
The larger relaxation time does not provide sufficient time for the elastic stresses to relax, and the surrounding medium behaves more like an elastic solid than a viscous liquid. 
Indeed, for sufficiently small bubble deformations, the natural frequency of a bubble in an elastic solid can be computed as  \citep{gaudron2015bubble,warnez2015numerical,dollet2019bubble}

\begin{equation}\label{eq:resfreq}
\omega_0 = \left\{\frac{1}{\rho R_0^2} \left[3k p_0 + 2(3k-1 )\frac{\gamma}{R_0} +4G\right] \right\}^{1/2}
\end{equation}
Thus, the elastic effects start affecting the resonance frequency when the shear modulus reaches values close to the atmospheric pressure $p_0 \approx 100$ kPa. 
Above this value, elasticity starts to dominate the response of the gas.
Even though Eq.~\eqref{eq:resfreq} is only valid for small amplitudes, it qualitatively shows how elasticity increases the resonance frequency. 
A fully quantitative confirmation of this scenario is provided in Fig.~\ref{fig:Oldroyd_B}. 
The figure shows that the resonance curve for a bubble inside an Oldroyd-B fluid with $G = 500$~kPa at finite $\lambda$ switches from a Newtonian liquid behavior ($\lambda\to 0$, blue dashed line)  to that of a Kelvin-Voigt solid ($\lambda \to \infty$, orange dashed line). 
Intriguingly, the transition is non-monotonic upon increasing $\lambda$, as is clearly demonstrated in Fig.~\ref{fig:Oldroyd_B}a.

As an additional perspective, we consider the resonance behavior of coated microbubbles.
Microbubbles can increase the contrast in ultrasound imaging or, once embedded in tissue, used as controlled cavitation agents for therapy. 
By contrast to the free gas bubble considered until now, these bubbles must be coated with a shell to prevent gas diffusion driven by the excess Laplace pressure.
The coating has two additional contributions to the bubble dynamics (see Appendix~\ref{app:coated}).
First, the dynamic change in surface area induces molecular friction, which leads to a significantly increased damping contribution \citep{van2007microbubble}. 
Second, the surface tension becomes a function of the packing density of the phospholipid molecules and, therefore, for a certain initial packing fraction, of the bubble size \citep{marmottant2005model}. 
When the bubble size decreases, the lipids get compressed leading to shell buckling like a spherical elastic shell. 
Using the viscoelastic Rayleigh-Plesset equation for coated microbubbles (see Appendix~\ref{app:coated}), we test the resonance curve of a coated microbubble inside an Oldroyd-B fluid with $\lambda = 1$ $\mu$s and $G = 500$ kPa.
We take a surface elasticity $\chi = 1$ N/m and a shell dilatational viscosity of $\kappa_s$ = $10^{-8}$ kg/s. 
The resulting resonance behavior of the coated microbubble is shown in Fig.~\ref{fig:Oldroyd_B}b. 
Compared to the result for the uncoated bubbles (Fig.~\ref{fig:Oldroyd_B}b), two features stand out. 
First, the resonances shift to slightly higher frequencies, which can be attributed to the additional stiffness induced by surface elasticity of the coating. 
Second, the peaks are wider, and of smaller amplitude, due to the additional damping induced by the shell's viscosity. 
Yet, the overall features of the resonance curves, specifically the trends with $\lambda$, remain unaffected by the coating, showing that these are dictated by the viscoelastic properties of the surrounding medium.

\begin{figure*}[t]
    \centering
    \includegraphics[width=0.975\linewidth]{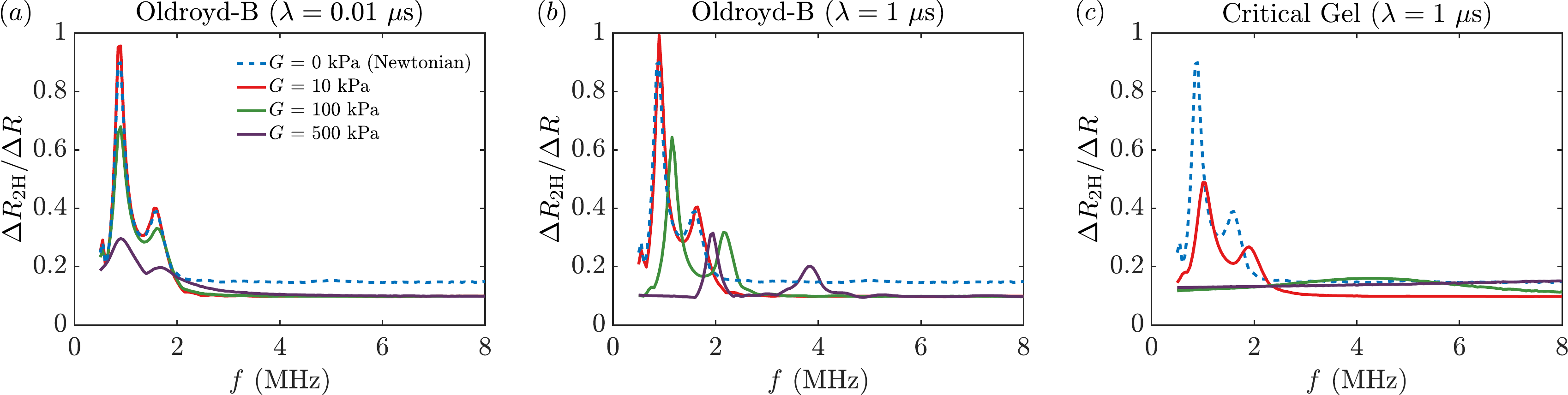}
    \caption{Resonance curve of the second harmonic response $\Delta R_{\mathrm{2H}}$ normalized by the fundamental amplitude $\Delta R$.
    (\textit{a}) For a relaxation time of $\lambda = 0.01$ $\mu$s, the trends of the second harmonics follow those of the fundamental.
    (\textit{b}) Similarly, increasing the relaxation time to $\lambda = 1$ $\mu$s we observe a shift of the resonance peak for higher shear moduli.
    Yet, at approximately half the resonance frequency a new peak appears.
    (\textit{c}) The similar trends between fundamental and second harmonic response can also be found for the Critical gel, where the amplitude of the second harmonic reduces by a factor of approximately 5.}
    \label{fig:Second}
\end{figure*}

\subsubsection{Critical Gel}

Polymer networks typically exhibit a broad spectrum of relaxation times, and therefore cannot be described by a simple exponential decay. 
For example, silicone gels are well-described by the so-called Chasset-Thirion model (cf. Table.~\ref{tab:constitutive}), which exhibit an algebraic decay in the range $0< n < 1$ \citep{LongAjdari1996,KarpitschkaNatureComm2015}. 
Here we focus on the special case of a material at the gelation point, for which the equilibrium shear modulus vanishes. Following Winter \& Chambon \citep{WinterChambon1986}, the Critical Gel corresponds to the case where $G'=G''$ at all frequencies (Fig.~\ref{fig:Ideal_Gel}d inset).
This criterion is achieved only for the special case of $\psi = G(\lambda/t)^{1/2}$, which gives  $G'=G''\sim G\lambda^{1/2} \omega^{1/2}$. Such a material at the gelation point has the peculiar property that, at low frequency, it has a vanishing shear modulus ($\lim_{\omega \to 0} G'= 0$) and an infinite steady viscosity ($\lim_{\omega \to 0} G''/\omega = \infty$). 
Conversely, at large frequency, the storage modulus diverges ($\lim_{\omega \to \infty} G'= \infty$) while the effective viscosity vanishes ($\lim_{\omega \to \infty} G''/\omega = 0$). 

The resonance behavior of free gas microbubbles inside a Critical Gel is indeed very different from that in an Oldroyd-B fluid. We find a stronger reduction in the amplitude of oscillation.
In addition, the Critical Gel also induces a much stronger resonance frequency shift.
This shift is not even visible for the lowest relaxation time of $\lambda = 0.01$ $\mu$s (Fig.~\ref{fig:Ideal_Gel}c), which did not induce a significant shift for the Oldroyd-B fluid. This is a direct consequence of the absence of a characteristic timescale in the Critical Gel. 
In fact, even though we separately defined $G$ and $\lambda$, the gel is entirely characterized by the single combination $G\lambda^{1/2}$, which is also known as the strength of the gel \citep{WinterChambon1986}. 
Therefore, the results for a larger relaxation time of $\lambda = 1$ $\mu$s (Fig.~\ref{fig:Ideal_Gel}d), shows similar trends as those for $\lambda = 0.01$ $\mu$s. 
More quantitatively, we can see in Fig.~\ref{fig:Ideal_Gel}c and d that the combination of [$G = 100$ kPa, $\lambda = 0.01$ $\mu s$] (green curve in Fig.~\ref{fig:Ideal_Gel}c) and [$G = 10$ kPa, $\lambda = 1$ $\mu s$] (red curve in Fig.~\ref{fig:Ideal_Gel}c) result in identical curves, since the product of $G\lambda^{1/2}$ is the equal.
Thus, for the gel, in contrast to the Oldroyd-B fluid, increasing $\lambda$ has a similar effect as increasing $G$.

The above results illustrate the importance of the rheology for the resonance behavior of microbubbles; in particular when comparing exponentially decaying relaxation functions to gel-like materials that exhibit a power-law spectrum. 
The fact that the strength of a critical gel involves the combination $G\lambda^{1/2}$ prevents the existence of an ``elastic limit" like for the Oldroyd-B fluid, since the equivalent solid would naturally have an infinite stiffness. 
The absence of an elastic limit at $\lambda \to \infty$ is not limited to the Critical Gel, but also applies to the Chasset-Thirion model that is used to describe silicone gels.

\subsubsection{Second harmonic}
Microbubbles driven at sufficiently high pressure amplitudes can undergo highly non-linear oscillations, generating higher order harmonic responses of the bubble \citep{prosperetti1974nonlinear,lauterborn1976numerical}, as well as subharmonic behavior \citep{neppiras1969subharmonic}.
In the context of viscoelastic fluids, Allen and Roy showed that the presence of elasticity can enhance the second harmonic response of an Upper Convected Maxwell fluid  \citep{allen2000dynamicsB}.
In this section we briefly comment on the second harmonic by examining the fast Fourier transform of the bubble response (Fig.~\ref{fig:R_vs_t}c). 
We record the oscillation amplitude $\Delta R_{\mathrm{2H}}$ at twice the driving frequency and directly compare the results with respect to the fundamental frequency in the context of the Oldroyd-B and Critical Gel of  Fig.~\ref{fig:Ideal_Gel}. 
The resonance curve of the second harmonic is shown in Fig.~\ref{fig:Second}. 
For the Oldroyd-B fluid with a relaxation time of $\lambda = 0.01$ $\mu$s we observe the appearance of two peaks for each curve (Fig.~\ref{fig:Second}a).
The main peak in the response occurs at approximately half the fundamental frequency, and a smaller one at the fundamental frequency.
The higher peak at half the fundamental frequency reflects the preference of the bubble to oscillate at its eigenfrequency.
An increase of the shear modulus dampens the curves and leads to smaller amplitudes.
The relaxation time is too short as compared to the oscillation period, and thus excites a viscous response from the Oldroyd-B fluid. 
Upon increasing the relaxation time to $\lambda = 1$ $\mu$s, the resonance frequency is shifted for the higher shear moduli  (Fig.~\ref{fig:Second}b).
The two peaks still appear at the fundamental and half the fundamental frequency but are shifted depending on the elasticity.
We thus find that the effects of the Oldroyd-B fluid on the second harmonic are similar to those observed for the fundamental frequency. 
The enhancement of the second harmonic response reported by Allen and Roy could therefore be attributed to the non-monotonic behavior of the Oldroyd-B model in the elastic limit.
Finally, the second harmonic of the Critical Gel also shares the same characteristics, as the trends follow the behavior of the fundamental frequency (Fig.~\ref{fig:Second}c). 
Two peaks also appear for certain values of $G\lambda^{1/2}$, with the amplitude decreasing as the shear modulus is increased.


\section{Summary and outlook}
In this article we have derived a unifying Rayleigh-Plesset-type equation for bubbles in viscoelastic materials.
Using concepts from finite linear viscoelasticity, we express the viscoelastic stresses as a function of the deformation history. This approach has the benefit of extending the Rayleigh-Plesset equation to a broad class of viscoelastic materials with arbitrary complex moduli through their relaxation function $\psi(t)$, while consistently accounting for large deformations.
Not only were we able to capture the bubble dynamics in materials with well-defined constitutive relations, such as the Upper Convected Maxwell fluid, but also to gel-like materials that can only be characterized by their storage and loss moduli, and not via a constitutive differential equation. 
With these ideas in mind, we tested the resonance behavior of a microbubble in different types of viscoelastic materials when driven by a tapered sinusoidal pressure pulse. 
The broad spectrum of relaxation times of gel-like materials completely alter the resonance behavior of a microbubble for sufficiently stiff shear moduli. 
On the other hand, the Oldroyd-B fluid was found to behave both as a viscous liquid and elastic solid, depending on the relaxation time. 
On the one hand, the Oldroyd-B fluid acts as a viscous liquid when the relaxation time is much faster than the typical oscillation period. 
Conversely, for relaxation times much slower than the typical oscillation period, the resonance curve of an Oldroyd-B fluid converges to that of a Kelvin-Voigt solid.

Our model is applicable to various viscoelastic materials; yet, it still has its limitations. Even though the bubble deformation is nonlinear in terms of the bubble radius $R$, our stress formulation of Eq.~\eqref{eq:KBKZfull} can be classified as finite linear viscoelastic, as the memory integral is still based on a superposition principle, involving a linear operator in $\mathbf{B}$. 
A possible extension to nonlinear response, while preserving the superposition principle, is by taking the relaxation function $\psi(t)$ to be dependent on the deformation \citep{wineman2009nonlinear}. 
For the derivation of Eq.\eqref{eq:RP_VE}, however, the assumption that $\psi(t)$is independent of deformation is crucial in our approach as it enabled us to separate the spatial and temporal integrals of the stress that lead to Eq.~\eqref{eq:tau_integration}. 
More generally, there are many constitute equations that are not described by Eq.~\eqref{eq:KBKZfull}. 
Typical examples include models that exhibit finite extensibility, which limits the extent to which the material can be stretched, such as the Gent model for solids \citep{gent1996new} or the FENE-P model for fluids \citep{warner1972kinetic}. 
Furthermore, viscoplastic materials also do not fall into the class of Eq.~\ref{eq:KBKZfull} \citep{de2019oscillations}.
The key advantage of Eq. \eqref{eq:RP_VE} therefore lies in the applicability to materials of arbitrary stress relaxation function, while preserving the possibility of finite deformations. 

Another limitation of our model is that we consider strictly radial bubble motions, i.e. purely spherical volumetric oscillations. 
This assumption can be consequential for sufficiently large oscillations, as the shape of a microbubble is susceptible to parametric instabilities when driven at large acoustic pressures \citep{versluis2010microbubble}.
The assumption of purely radial motions is also crucial when extending our model to coated microbubbles, which are known to exhibit non-spherical oscillations either as a result of a symmetry breaking in the medium, or as surface modes develop \citep{dollet2008nonspherical,dollet2019bubble}. 
As a future perspective, it would thus be worthwhile focusing on the non-spherical motion of coated and uncoated microbubbles in viscoelastic materials. 
Indeed, our theoretical approach lays a generalized framework for bubble dynamics in viscoelastic media, which, as we have exemplified, could also be extended to model  viscoelastic dissipation at the interface of coated microbubbles.



\begin{acknowledgments}
\end{acknowledgments}

\appendix
\section{Lower convected derivative}
\label{app:Lower}
Our analysis focused on upper convected materials by setting $\psi_2 = 0$ in Eq.~\eqref{eq:KBKZfull}.
If we instead had chosen to work with the lower convected derivative, it would require setting $\psi_1 = 0$ and using the inverse of $\mathbf{B}(t,t')$.
The analysis is still straightforward, owing to the purely radial flow that leads to a diagonal Finger tensor.
The components of the deviatoric stress tensor then adopt the form
\begin{equation}
\begin{split}
    \tau_{rr} = 4\int_{-\infty}^{t}\mathrm{d}t'\,&\{\psi_2(t-t') R^2(t')\,\dot{R}(t')\\
    & \frac{\left[\mathcal{R}^3+R^3(t)-R_0^3\right]^{4/3}}{\left[\mathcal{R}^3+R^3(t')-R_0^3\right]^{7/3}}\},
\end{split}
    \label{eq:tau_rr_Lower}
\end{equation}
\begin{equation}
\begin{split}
    \tau_{\theta\theta} = \tau_{\phi\phi} = -2\int_{-\infty}^{t}\mathrm{d}t'\,&\{\psi_2(t-t')R^2(t')\,\dot{R}(t')\\
    & \frac{\left[\mathcal{R}^3+R^3(t')-R_0^3\right]^{-1/3} }{\left[\mathcal{R}^3+R^3(t)-R_0^3\right]^{2/3}}\}.
\end{split}
    \label{eq:tau_phi_Lower}
\end{equation}
The radial integration of the stresses in the Rayleigh-Plesset equation again leaves only the temporal memory integrals such that
\begin{equation}
\begin{split}
    &\int_{R}^{\infty}\mathrm{d}r\,\frac{1}{r} (2\tau_{rr} - \tau_{\theta\theta}- \tau_{\phi\phi}) = \\
    &- 2\int_{-\infty}^{t}\mathrm{d}t'\,\psi(t-t')\,\frac{\dot{R}(t')}{R(t)}\left[\frac{R(t)}{R(t')} \right]^2\left\{  
     \left[\frac{R(t')}{R(t)} \right]^3 +   1\right\}.
\end{split}
\end{equation}
Using the lower convected formulation leads to an additional factor $[R(t)/R(t')]^2$ in the memory integral.

\section{Kinematics}
\label{app:Kinematics}
Here we provide an overview of the kinematic relations that can be used to better interpret the stress Eq.~\eqref{eq:KBKZ} in certain limits.
Recall that the deformation gradient tensor is defined as $\mathbf{F}(t,t') = \frac{\partial \mathbf{x}}{\mathbf{\partial x'}}$.
Taking the time derivative $\mathrm{d}/\mathrm{d}t$ at a constant material point and using the chain rule, one can show that \citep{morozov2015introduction,snoeijer2020relationship,essink2022soft,stone2023note}
\begin{subequations}
\begin{gather}
	\frac{\mathrm{d} \mathbf{F}}{\mathrm{d} t} = (\mathbf{\nabla}\mathbf{v})^{\mathrm{T}} \cdot \mathbf{F} \\
	\frac{\mathrm{d} \mathbf{F}^{-1}}{\mathrm{d} t} = -\mathbf{F}^{-1}\cdot(\mathbf{\nabla}\mathbf{v})^{\mathrm{T}}
\end{gather}
	\label{eq:dFdt}
\end{subequations}
where $\mathbf v(\mathbf x)$ is the Eulerian velocity field. 

Considering now the deformations with respect to past time $t'$, passing via the reference state at $t_0$, the deformation tensor can be expressed as $\mathbf{F}(t,t') = \mathbf{F}(t)\cdot\mathbf{F}^{-1}(t')$, where for notational convenience we use a single argument when the reference state is involved, i.e. $\mathbf F(t)=\mathbf F(t,t_0)$.
As a result, the Finger tensor becomes
\begin{equation}
\begin{split}
	\mathbf{B}(t,t') & = \mathbf{F}(t,t')\cdot\mathbf{F}^{\mathrm{T}}(t,t')\\
	 &=  \mathbf{F}(t)\cdot\mathbf{F}^{-1}(t')\cdot\mathbf{F}^{-\mathrm{T}}(t')\cdot\mathbf{F}^{\mathrm{T}}(t),
\end{split}
\end{equation}
and its inverse
\begin{equation}
\begin{split}
	\mathbf{B}^{-1}(t,t') & = \mathbf{F}^{-\mathrm{T}}(t,t')\cdot\mathbf{F}^{-1}(t,t')\\
	 &= \mathbf{F}^{-\mathrm{T}}(t)\cdot\mathbf{F}^{\mathrm{T}}(t')\cdot\mathbf{F}(t')\cdot\mathbf{F}^{-1}(t).
\end{split}
\end{equation}
We now wish to take the time derivative of $\mathbf{B}(t,t')$ with respect to $t'$ as expressed in the stress in Eq.~\eqref{eq:KBKZ}.
Making use of the kinematic relations Eq.~\eqref{eq:dFdt}, we obtain  \citep{essink2022soft}
\begin{equation}
\begin{split}
&\frac{\partial \mathbf{B}(t,t')}{\partial t'} = \\
&  \mathbf{F}(t)\cdot \left[\frac{\mathrm{d} \mathbf{F}^{-1}(t')}{\mathrm{d}t'}\cdot\mathbf{F}^{-\mathrm{T}}(t') + \mathbf{F}^{-1}(t') \cdot \frac{\mathrm{d} \mathbf{F}^{-\mathrm{T}}(t')}{\mathrm{d}t'} \right]\cdot\mathbf{F}^{\mathrm{T}}(t) \\
& =-\mathbf{F}(t)\cdot \mathbf{F}^{-1}(t')\cdot\left[(\mathbf{\nabla}\mathbf{v})^{\mathrm{T}}(t') + \mathbf{\nabla}\mathbf{v}(t')\right]\cdot\mathbf{F}^{\mathrm{-T}}(t') \cdot\mathbf{F}^{\mathrm{T}}(t) \\
& =-\mathbf{F}(t,t')\cdot \dot{\boldsymbol{\epsilon}}(t')\cdot\mathbf{F}^{\mathrm{T}}(t),
\end{split}
\end{equation}
where we recognize the rate of strain tensor $\dot{\boldsymbol\epsilon} = (\mathbf{\nabla}\mathbf{v})^{\mathrm{T}} + \mathbf{\nabla}\mathbf{v}$. This proves that, in general, one cannot replace $\frac{\partial \mathbf{B}(t,t')}{\partial t'}$ by $\dot{\boldsymbol\epsilon}$ in the constitutive relation. 
Similarly for the inverse of the Finger tensor we get \citep{essink2022soft}
\begin{equation}
\frac{\partial \mathbf{B}^{-1}(t,t')}{\partial t'} = \mathbf{F}^{-\mathrm{T}}(t,t')\cdot \dot{\boldsymbol{\epsilon}}(t')\cdot\mathbf{F}^{-1}(t,t').
\end{equation}
Only when taking the limit $t'\to t$, or when deformations are small, we recover by definition that $\mathbf{F}(t,t') \simeq \mathbf{I}$, and we observe that the time derivatives of $\mathbf{B}(t,t')$ and its inverse reduce to $\mp\dot{\boldsymbol \epsilon}(t)$.

\section{Convected Maxwell Models}
\label{app:UCM}
We show how, for special choices of $\psi(t)$, the integral formulation for the stress can be expressed as a constitutive differential equations (Tab.~\ref{tab:constitutive}).
We first introduce the upper convected derivative 
\begin{equation}
\overset{\triangledown}{\mathbf{A}} = \frac{\mathrm{d} \mathbf{A}}{\mathrm{d}t} - (\mathbf{\nabla}\mathbf{v})^{\mathrm{T}}\cdot \mathbf{A} - \mathbf{A}\cdot \mathbf{\nabla}\mathbf{v},
\end{equation}
which ensures that the dynamics of a second rank tensor $\mathbf{A}$ remain frame invariant.
The first term is a time derivative evaluated at a constant material point, while the last two terms ensure that $\mathbf{A}$ transforms appropriately as it gets deformed by the flow. 
Applying the upper convected derivative to Eq.~\eqref{eq:KBKZ}, we get
\begin{equation}
\begin{split}
\overset{\triangledown}{\boldsymbol\tau} & =  
\frac{\mathrm{d} \boldsymbol\tau}{\mathrm{d}t} - (\mathbf{\nabla}\mathbf{v})^{\mathrm{T}}\cdot \boldsymbol\tau - \boldsymbol\tau\cdot \mathbf{\nabla}\mathbf{v} \\
& = -\left.\psi(t-t') \frac{\partial \mathbf{B}(t,t')}{\partial t'}\right|_{t'=t} - (\mathbf{\nabla}\mathbf{v})^{\mathrm{T}}\cdot \boldsymbol\tau - \boldsymbol\tau\cdot \mathbf{\nabla}\mathbf{v}  \\
&-\int_{-\infty}^t\mathrm{d}t'\,\left[\frac{\mathrm{d}\psi(t-t')}{\mathrm{d}t} \frac{\partial \mathbf{B}(t,t')}{\partial t'}+ \psi(t-t') \frac{\partial^2 \mathbf{B}(t,t')}{\partial t\partial t'}\right]\\
& = \psi(0)\dot{\boldsymbol{\epsilon}}(t) -\int_{-\infty}^t\mathrm{d}t'\,\frac{\mathrm{d}\psi(t-t')}{\mathrm{d}t} \frac{\partial \mathbf{B}(t,t')}{\partial t'}.
\end{split}
\end{equation}
Note that we have utilized the relation  $\frac{\partial^2 \mathbf{B}(t,t')}{\partial t\partial t'} = (\mathbf{\nabla}\mathbf{v})^{\mathrm{T}}(t)\cdot \frac{\partial \mathbf{B}(t,t')}{\partial t'} + \frac{\partial \mathbf{B}(t,t')}{\partial t'}\cdot \mathbf{\nabla}\mathbf{v}(t)$, which exactly cancels the last two terms from the upper convected derivative.

Upon taking the upper convected derivative, we thus transformed the integral equation to a integro-differential equation. So, not much progress is made, except for special choices for the stress relaxation function. Specifically, we recover the Upper Convected Maxwell model for $\psi(t) = G\exp(-t/\lambda)$. Using this relaxation function, we obtain the differential equation
\begin{equation}
\boldsymbol\tau + \lambda \overset{\triangledown}{\boldsymbol\tau} = \eta \dot{\boldsymbol\epsilon},
\end{equation}
where we have introduced the Maxwell viscosity $\eta = G\lambda$.
Indeed, this equation is the conventional form of the Upper Convected Maxwell model for large deformations \citep{bird1987dynamics}, which was also used by Allen and Roy \citep{allen2000dynamicsA} to study large amplitude bubble oscillations in viscoelastic liquids.

For completeness, we also apply the same analysis for the Lower Convected Maxwell Model.
We first introduce the lower convected derivative 
\begin{equation}
\overset{\triangle}{\mathbf{A}} = \frac{\mathrm{d} \mathbf{A}}{\mathrm{d}t} + (\mathbf{\nabla}\mathbf{v})\cdot \mathbf{A} + \mathbf{A}\cdot (\mathbf{\nabla}\mathbf{v})^{\mathrm{T}},
\end{equation}
which has the same property of preserving frame invariance.
We set $\psi_1(t) = 0$ in Eq.~\eqref{eq:KBKZfull} and retain the $\psi_2(t)$ relaxation function.
Applying the lower convected derivative to the stress we get
\begin{equation}
\begin{split}
\overset{\triangle}{\boldsymbol\tau} & =  
\frac{\mathrm{d} \boldsymbol\tau}{\mathrm{d}t} +  (\mathbf{\nabla}\mathbf{v})\cdot \boldsymbol\tau + \boldsymbol\tau\cdot (\mathbf{\nabla}\mathbf{v})^{\mathrm{T}} \\
& = \left.\psi_2(t-t') \frac{\partial \mathbf{B}^{-1}(t,t')}{\partial t'}\right|_{t'=t} +  (\mathbf{\nabla}\mathbf{v})\cdot \boldsymbol\tau + \boldsymbol\tau\cdot (\mathbf{\nabla}\mathbf{v})^{\mathrm{T}}  \\
&+\int_{-\infty}^t\mathrm{d}t'\,\left[\frac{\mathrm{d}\psi_2(t-t')}{\mathrm{d}t} \frac{\partial \mathbf{B}^{-1}(t,t')}{\partial t'}+ \psi_2(t-t') \frac{\partial^2 \mathbf{B}^{-1}(t,t')}{\partial t\partial t'}\right] \\
& = \psi_2(0)\dot{\boldsymbol{\epsilon}}(t) +\int_{-\infty}^t\mathrm{d}t'\,\frac{\mathrm{d}\psi_2(t-t')}{\mathrm{d}t} \frac{\partial \mathbf{B}^{-1}(t,t')}{\partial t'}.
\end{split}
\end{equation}
Similar to the upper convected case, the relation $\frac{\partial^2 \mathbf{B}^{-1}(t,t')}{\partial t\partial t'} = -\mathbf{\nabla}\mathbf{v}(t)\cdot \frac{\partial \mathbf{B}^{-1}(t,t')}{\partial t'} - \frac{\partial \mathbf{B}^{-1}(t,t')}{\partial t'}\cdot (\mathbf{\nabla}\mathbf{v})^{\mathrm{T}}(t)$ yields exact cancellations with the last two terms of the lower convected derivative.
When we apply the relaxation function of the lower convected derivative $\psi_2(t) = G\exp{(-t/\lambda)}$, we get the standard form of the Lower Convected Maxwell model
\begin{equation}
\boldsymbol\tau + \lambda \overset{\triangle}{\boldsymbol\tau} = \eta \dot{\boldsymbol\epsilon}.
\end{equation}

\section{Coated microbubbles}
\label{app:coated}
We show how the Rayleigh-Plesset equation gets modified when considering the dynamics of coated microbubbles.
Introducing a coating around the bubble introduces an additional dilatational viscosity $\kappa_s$ and a surface elasticity $\chi$.
The effect of the surface elasticity is well-described by the Marmottant model, which consists of a piece-wise function for the surface tension $\gamma(R)$ \citep{marmottant2005model}:
\begin{equation*}
\gamma(R)=\begin{cases}
          0\quad &\text{if} \,\,R \leq R_{\mathrm{buck}} \\
          \gamma(R_0)+\chi\left(\frac{R^2}{R_{\mathrm{buck}}^2}-1 \right) \quad &\text{if} \,\, R_{\mathrm{buck}} \leq R \leq R_{\mathrm{rup}} \\
          \gamma_l \quad &\text{if}\, R \geq R_{\mathrm{rup}} \\
     \end{cases}
\end{equation*}
Here, $R_{\mathrm{buck}}$ denotes the radius at which the coating buckles and below which the surface tension is zero, $R_{\mathrm{rup}}$ the radius at which the coating ruptures.
Above this radius, the free gas interface is exposed to the surrounding medium and the surface tension is equal to that of the medium $\gamma_l$.
The parameter $\chi$ denotes the shell elasticity of the coating, i.e. the rate of change in surface tension with respect to the bubble surface area $A$: $\chi = A \frac{\mathrm{d} \gamma}{\mathrm{d}A}$.
Molecular dissipation is introduced through a dilatational viscosity $\kappa_s$, associated with the phospholipid monolayer.
As a result, the Rayleigh-Plesset equation now becomes

\begin{equation}
\begin{split}
    & \rho \left(R \Ddot{R} + \frac{3}{2}\dot{R}^2\right) = \left(p_0 + \frac{\gamma(R_0)}{R_0} \right) \left(\frac{R_0}{R} \right)^{3k} - \frac{2\gamma(R)}{R} -p_\infty\\ 
    &- \frac{4\kappa_s\dot{R}}{R^2}- 2\int_{-\infty}^{t}\mathrm{d}t'\,\psi(t-t')\,\frac{\dot{R}(t')}{R(t)}\left\{  
     \left[\frac{R(t')}{R(t)} \right]^3 +   1 \right\}.
\end{split}
         \label{eq:RP_Coated}
\end{equation}











\end{document}